\pdfoutput = 1
\documentclass[apj,numberedappendix]{emulateapj}

\usepackage{epsfig}
\usepackage{amsmath}
\usepackage{enumerate}
\usepackage{mathrsfs}

\usepackage{natbib}
\usepackage{hyperref}

\usepackage{ulem}
\usepackage[squaren, Gray, cdot]{SIunits}

\usepackage{color}

\newbox\grsign \setbox\grsign=\hbox{$>$} \newdimen\grdimen \grdimen=\ht\grsign
\newbox\simlessbox \newbox\simgreatbox \newbox\simpropbox
\setbox\simgreatbox=\hbox{\raise.5ex\hbox{$>$}\llap
     {\lower.5ex\hbox{$\sim$}}}\ht1=\grdimen\dp1=0pt
\setbox\simlessbox=\hbox{\raise.5ex\hbox{$<$}\llap
     {\lower.5ex\hbox{$\sim$}}}\ht2=\grdimen\dp2=0pt
\setbox\simpropbox=\hbox{\raise.5ex\hbox{$\propto$}\llap
     {\lower.5ex\hbox{$\sim$}}}\ht2=\grdimen\dp2=0pt

\def\beq{\begin{equation}}
\def\eeq{\end{equation}}

\begin{document}

\title{Resonant friction on discs in galactic nuclei}

\author{Yuri Levin$^{1,2,3}$.
}

\affil{$^1$Physics Department and Columbia Astrophysics Laboratory, Columbia University, 
538 West 120th Street, New York, NY 10027}
\affil{$^2$Center for Computational Astrophysics, Flatiron Institute, 162 5th Ave, NY10011}
\affil{$^3$School of Physics and Astronomy, Monash University, Clayton VIC 3800, Australia}

\begin{abstract}
We argue that resonant friction has a dramatic effect on a disc 
whose rotation direction is misaligned with that of its host nuclear star cluster. The disc's gravity causes gravitational perturbation of the cluster that in turn exerts a strong torque back onto the disc. We argue that this torque may be responsible for the observed disruption of the clockwise disc of young stars in the Galactic Center, and show in numerical experiments that it produces the observed features in the distribution of the stars' angular momenta. More generally, we speculate that the rotation of nuclear star clusters has a stabilizing effect on the orientation of transient massive accretion discs around the supermassive black holes residing in their centers, and thus on the directions and magnitudes of the black-hole  spins.


\medskip
\end{abstract}

\keywords{}

\section{introduction}
About $100$ young massive stars are located in the central half-parsec of the Milky-Way's nuclear star cluster, with a supermassive black hole at is center. The proper motions of these stars indicate an average sense of clockwise rotation  around the black hole \citep{2000MNRAS.317..348G}. More detailed 3-dimensional analyses show that a significant fraction ($\sim 30-60 \%$) of these stars are moving in a thin clockwise disc \citep{2003ApJ...590L..33L, 2003ApJ...594..812G, 2006ApJ...643.1011P, 2006ApJ...648..405B, 2009ApJ...690.1463L,  2009ApJ...697.1741B, 2014ApJ...783..131Y, 2022ApJ...932L...6V}. The other stars clearly do not belong to the disc, with several groups disagreeing about the interpretation of their precise geometry. \cite{2003ApJ...594..812G}, \cite{2006ApJ...643.1011P}, \cite{2009ApJ...697.1741B}, and \cite{2022ApJ...932L...6V}   argue that the other stars form another disc-like structure, albeit with somewhat diffuse clustering of the stars' orbital planes, possibly connected by a strong warp to the clockwise disc.  \cite{2022ApJ...932L...6V} goes further and claims that several diffuse disc-like structures can be identified. By contrast, \cite{2009ApJ...690.1463L} and \cite{2014ApJ...783..131Y} argue that all these secondary disc-like structures are not statistically significant. These groups also disagree on the precise fraction of the young stars that belong to the clockwise disc.

It is appealing to consider this kinematic data as a result of a partial disruption of an initially coherent stellar disc. There is strong evidence that the young stars were formed {\it in situ} \citep{2005MNRAS.364L..23N, 2006ApJ...643.1011P, 2009ApJ...690.1463L}, and by far the most natural scenario for this is the star formation inside a gravitationally unstable gas disc \citep{2003ApJ...590L..33L, 2008Sci...321.1060B} . \cite{2009A&A...496..695S} suggested that a torque from the circumnuclear gas ring located at a distance $\sim 2$pc from the black hole, could exert a disruptive torque on the disc, but their analysis did not take the disc's self-gravity into account. A direct $N$-body simulations of this process by \cite{2016ApJ...818...29T} showed that it was not efficient in disrupting the disc. \cite{2011MNRAS.412..187K, 2015MNRAS.448.3265K, 2022MNRAS.tmp.2848P} explored whether the disc could be disrupted by stochastic torques from vector resonant relaxation (VRR), but found that in order for this to happen in $5\times 10^6$ years, the masses of the background stars in the cluster had to be unrealistically high, over $100M_{\odot}$.

We think that the key to the puzzle of the disrupted disc is in its gravitational interaction with the rotating nuclear star cluster inside which it resides. More specifically, we show that the effect called ``resonant friction'' 
can produce a very strong torque on the disc that tends to align the disc's rotation with that of the cluster, and in the process can partially disrupt 
the disc in less that $5\times 10^6$ years. The Milky-way's nuclear cluster is rotating neither clockwise nor counterclockwise relative to the line of sight; instead its rotational axis is close to that of the Galaxy, and thus the cluster's rotation is misaligned with that of the disc \citep{2014A&A...570A...2F}. In Section 2 we demonstrate using numerical simulations that this configuration naturally leads to the disc disruption, and  that the expected orbital distribution is qualitatively similar to the one that is observed in the Galactic Center. Typically we see an inner disc, sometimes warped or broken up into rings, co-existing with a more diffuse disoriented or weakly clustered orbits of the outer stars, with the inner disc containing $\sim 50\%$ of the total stars. The reader uninterested in the details of this paper should just look at Figure 1: it contains the paper's most interesting results. In this section we also explain how we choose the range of rotational parameters of the cluster based on existing observations of the cluster's mean radial velocities. In Section 3 we back up the numerical results by obtaining  an analytical estimate of  the Resonant Friction timescale inside a slowly-rotating cluster. In Section 4 we speculate that resonant friction would reorient gaseous accretion discs that are  formed inside a nuclear star cluster from radially infalling, tidally captured gas clouds. We discuss what impact this would have on spins of supermassive black holes. We present our numerical algorithm in the Appendix.
\begin{figure*}
\centering
   
   \includegraphics[width=.450\textwidth]{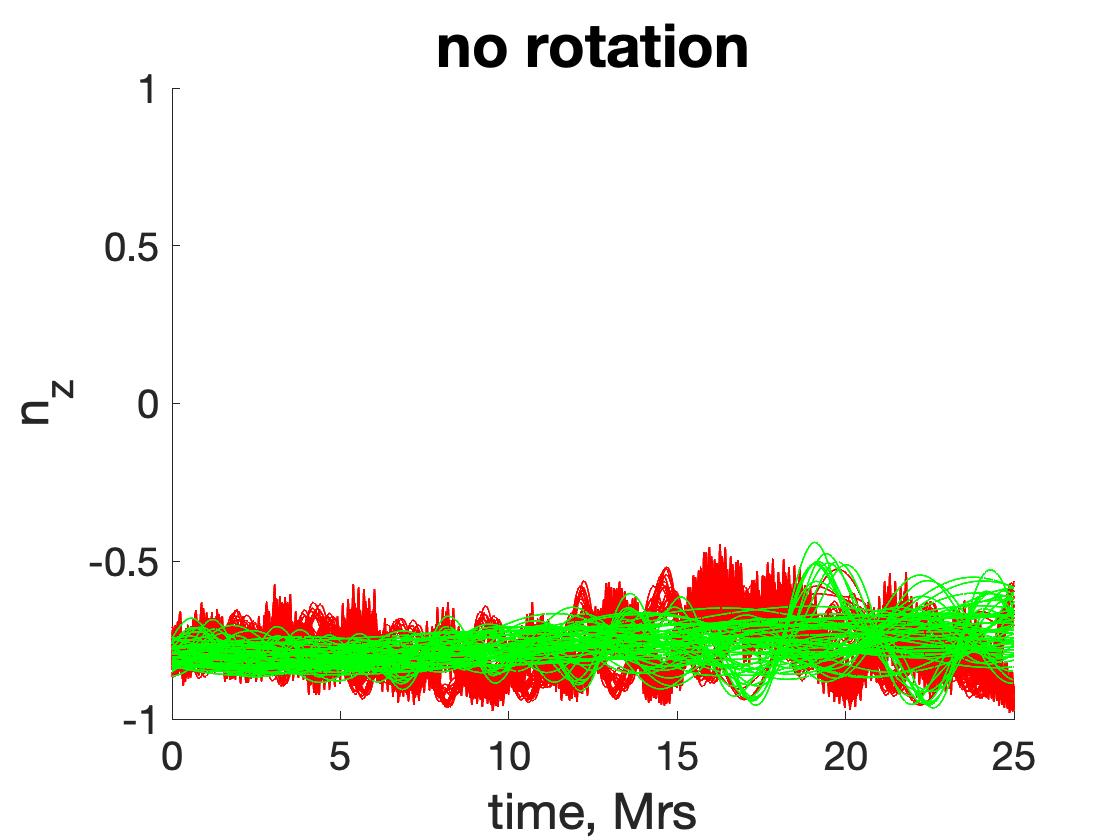}
   \includegraphics[width=.450\textwidth]{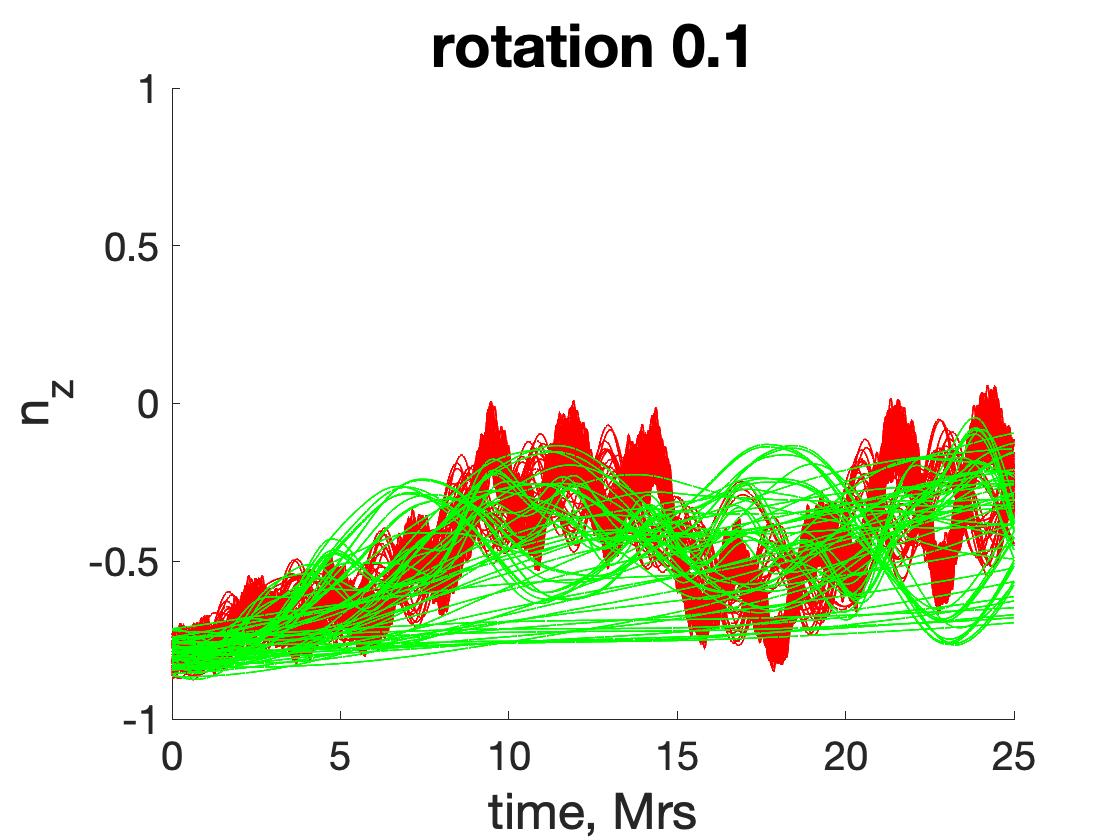} 
   \includegraphics[width=.450\textwidth]{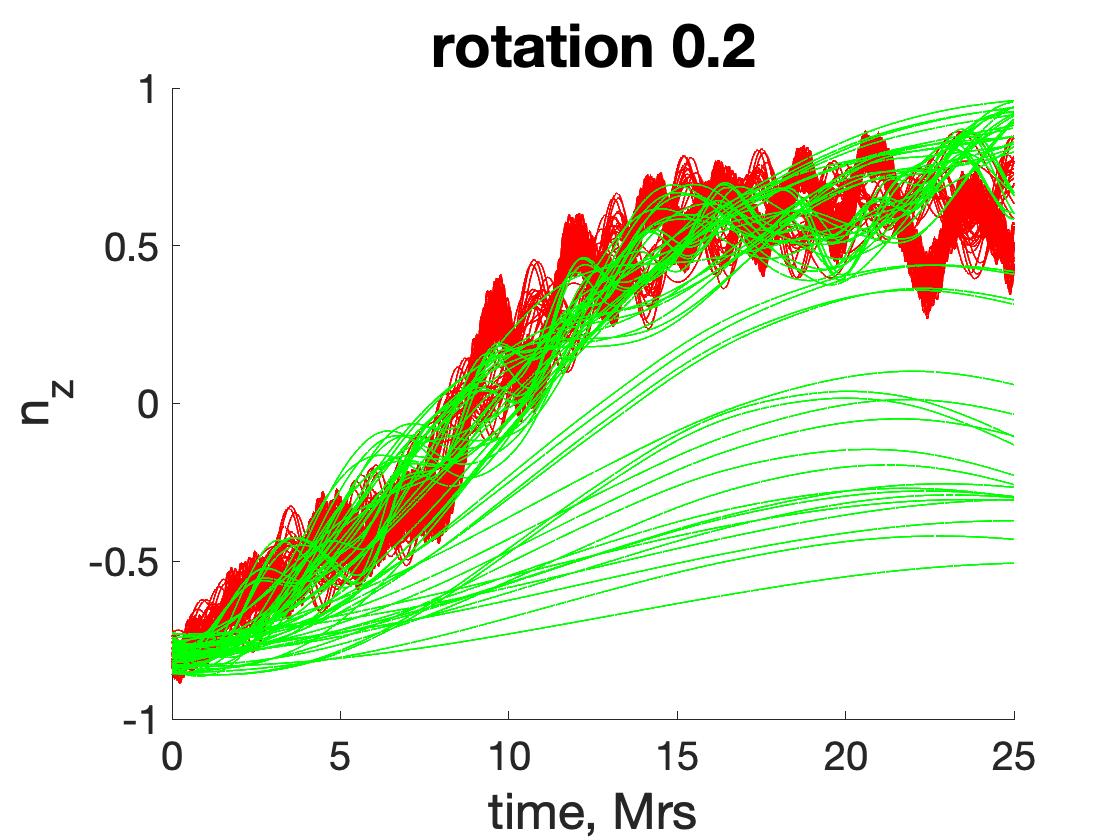}
   \includegraphics[width=.450\textwidth]{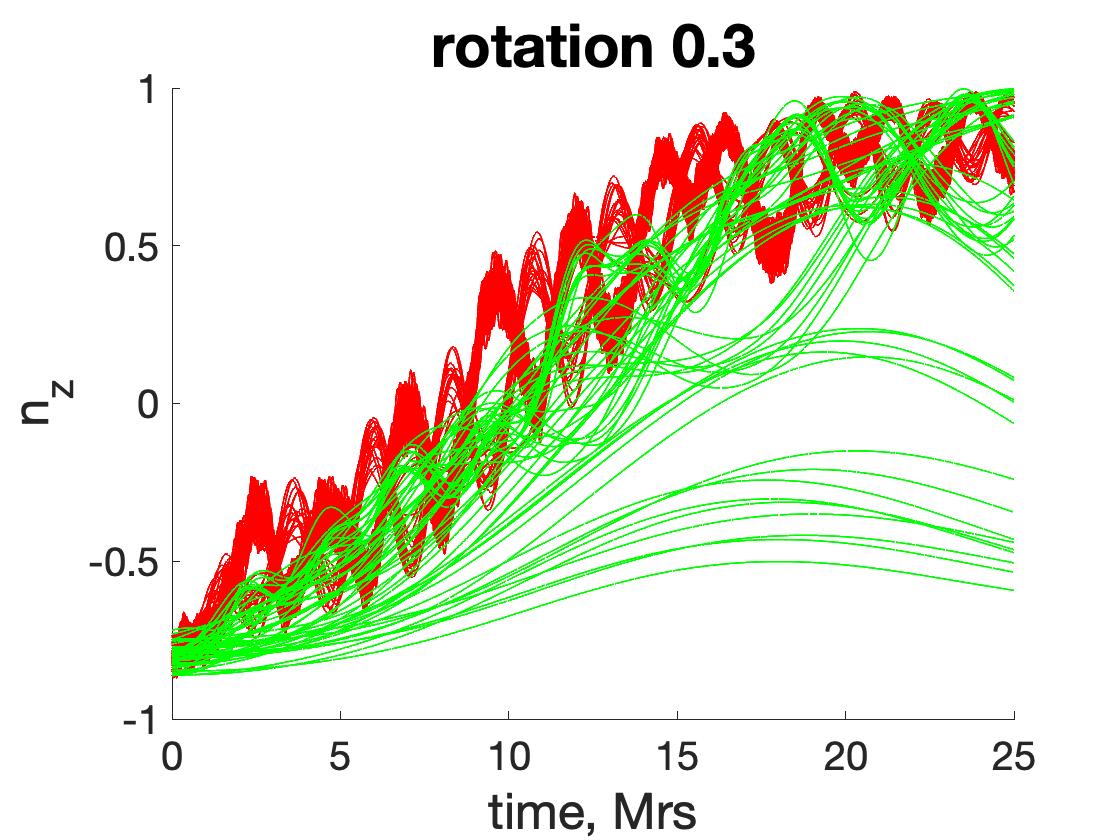}
   \includegraphics[width=.45\textwidth]{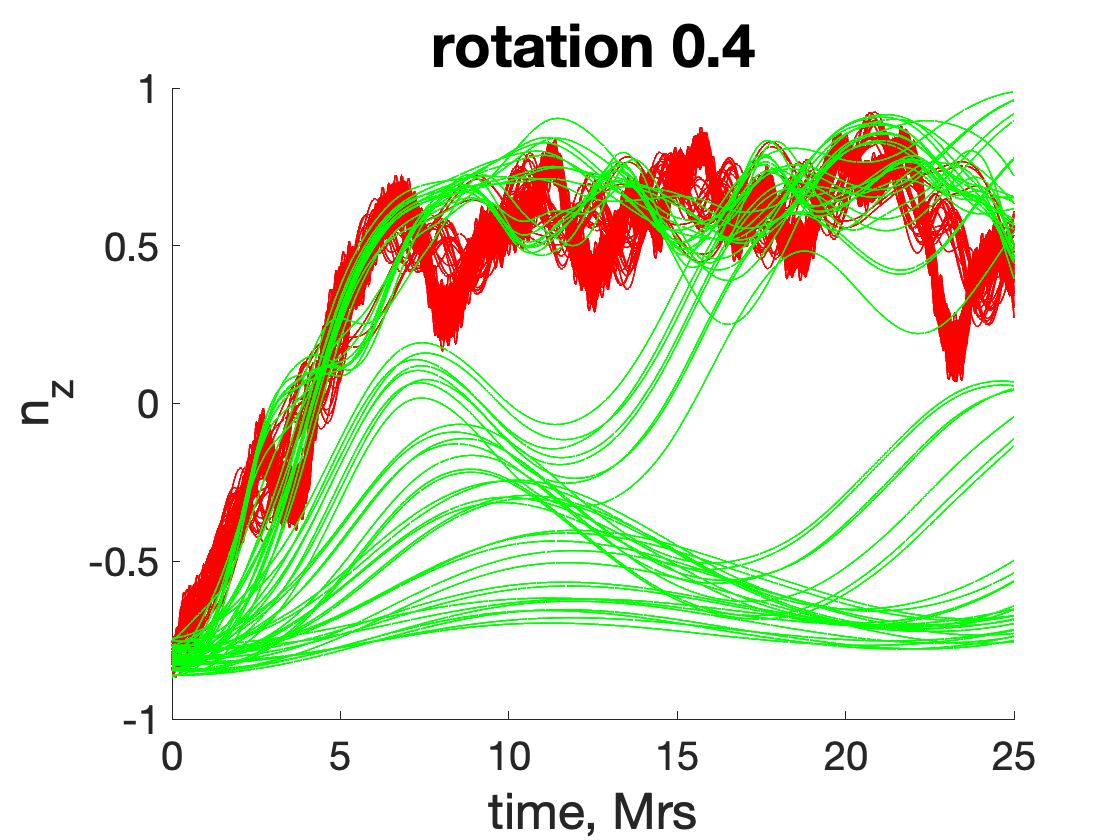}
   \includegraphics[width=.45\textwidth]{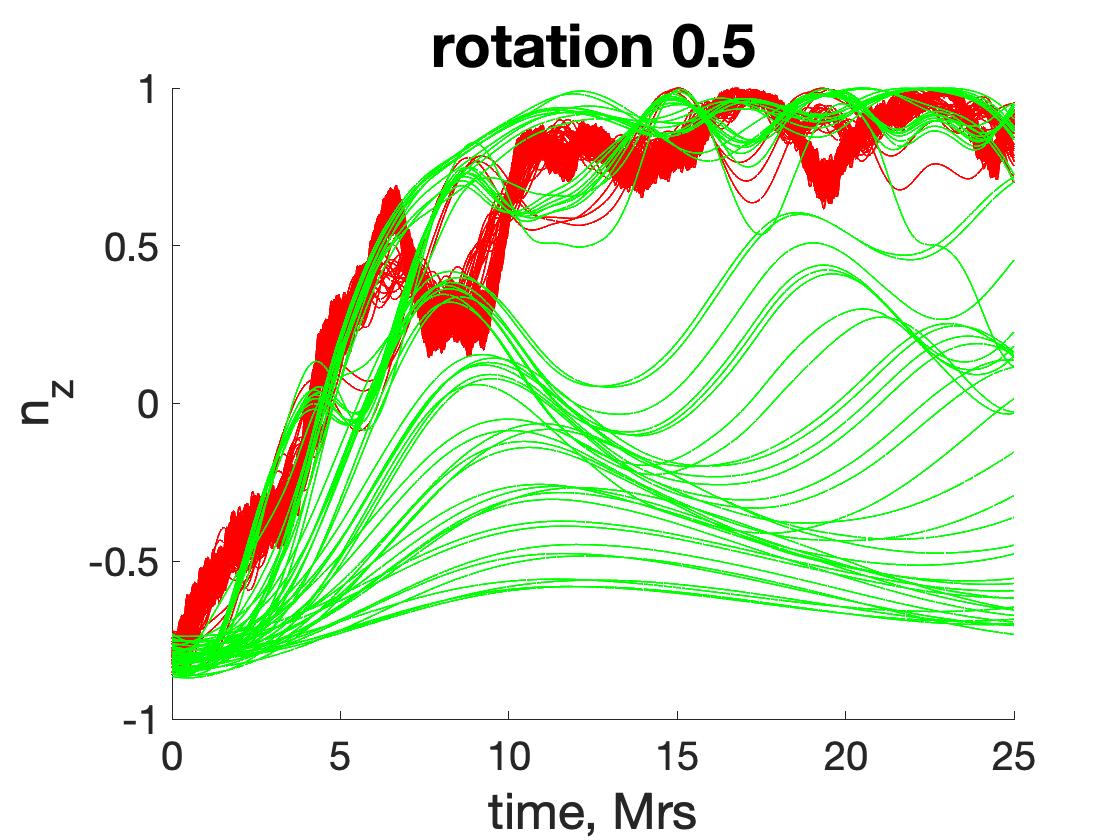}
 \caption{{\bf Evolution of the stellar disc.} The background cluster is initially rotating along the $z$ axis with the dimensionless rotation rate $\tilde{\gamma}$ defined in the text and specified at the top of each subfigure. The counter-rotating disc is injected at $t=0$. The figure shows some randomly chosen examples of the time evolution of $n_z$, where $\vec{n}$ is the unit vector directed along the star's angular momentum,  for all the stars in the disc. The red lines are marking the inner half of the disc's stars and the green lines are marking those the outer half. As the rotation rate increases, we observe the disruption of the outer parts of the disc and tendency for the inner disc to align with the cluster. At high rotation rate, the disc gets warped or  broken into transient rings. These effects are very important at $5$Mys, the likely age of the young stellar population in the Galactic Center.
 \small
 }
\end{figure*}
\newline\newline
\section{The stellar disc inside a rotating cluster}
\subsection{The results of numerical experiments}
We model only the inner part of the central cluster, which in our simulations consists of $1.2\times 10^5$ solar-mass stars, located inside the central $0.4$pc with the spacial number density of stars given by the Peebles-Young distribution,  $n(r)\propto r^{-1.5}$. The cluster, if it were extended out to $1$pc, would contain about half a million solar masses. The numbers are  based on \cite{2007A&A...469..125S}, but it is questionable how faithfully they represent the distribution of dynamical scatterers in the Galactic Center, since (a) there seems to be a hole in the stellar distribution in the central $\sim 0.2$pc \citep{2009A&A...499..483B, 2009ApJ...703.1323D, 2010ApJ...708..834B}, and (b) stellar-mass black holes are expeceted to segregate inside central $0.1$pc \citep{2000ApJ...545..847M, 2009ApJ...697.1861A}.

Despite these uncertainties, we believe that  the numbers are correct to within an order of magnitude. Our young massive disc consists of $105$  stars  of $80 M_\odot$ each. It is the total mass of the disc that plays key role, and we get similar results if we increase the number of stars and decrease their masses, while keeping the total mass constant. The disc's surface density is $\Sigma\propto r^{-2}$ \citep{2006ApJ...643.1011P} and its outer edge is at $0.4$pc.

In our numerical experiments, we sacrifice the precision of individual interactions between stellar orbits in favour of extremely rapid speed and ease with which we can explore physical effects. The fastest dynamics responsible for the time evolution in the orientation of  orbital planes, is captured when the precessing orbits are replaced with the annuli that they trace; this type of relaxation is called ``vector resonant relaxation" \citep{1996NewA....1..149R}. We however, replace each annulus with a circular ring of comparable radius, and  keep only quadrupolar and octupolar terms in the interaction potential between the rings. Furthermore, instead of considering a continuum of the ring radii, we choose a modest number of discreet radius values, i.e., ``the bins" ($10$ for the computations shown in Figure 1). Our computational vectorization benefits from each bin having an equal number of stars, including the stars that belong to the disc; there is no restriction on the stars' individual masses. The requirement that the overall distribution (approximately) represents $n(r)\propto r^{-1.5}$  determines the value of the radius assigned to each bin. The main advantage of the binning is that it allows us to perform $\sim N_{s}$ instead of $\sim N_s^2$ operations per each step. This point, together with other details of our numerical procedure, is explained in the Appendix. The commented Matlab code is available upon request.

Each simulation runs models $3\times 10^7$ years of evolution, on a usual desktop computer. It takes several minutes to complete if only quadrupolar terms are taken into account, and several hour if the octupole terms are included.
We use the "octupolar" simulations for the production runs, and the ``quadrupolar'' ones as a super-fast way to explore the parameter space. The only significant difference is that the inner half of the disc tends to remain somewhat more coherent if the octupolar terms are included. This allows us to quickly change the parameters of the system, and explore the convergence and stability of the results. The results of the numerical simulations are illustrated in Figure 1. The disc is injected in counter-rotation with respect to the cluster, so that the $\vec{n}_i=\vec{l}_i/l_i$ are clustered in the southern hemisphere relative to the cluster rotation; here $\vec{l}_i$ is the angular momentum of the $i$'th star. If the cluster is non-rotating, the disc stays coherent even though its bending modes are excited, consistent with the findings of \cite{2011MNRAS.412..187K,2015MNRAS.448.3265K} and, most convincingly, with the direct simulations of \cite{2022MNRAS.tmp.2848P}. For greater rotation rates, we see that within several million years, the  orbits of the inner half of the stars (represented by the red lines) are dragged towards corotation with the cluster, while the outer stars get dispersed. For more rapidly rotating clusters we observe the disc being warped or split into distinctly oriented rings, which are seen as coherent groups of the $\vec{n}$-vectors. This complexity is consistent with the phenomenology seen in the galactic center.

What is the rotation rate of the cluster?
In the next subsection we mathematically define the dimensionless rotation rate, and show how to estimate it from the radial velocity data. Using existing literature, we obtain a  rough estimate of $\tilde{\gamma}\sim 0.3$.

\subsection{The rotation rate of the relaxed cluster}
 Inside the gravitational radius of influence of a supermassive black hole, the orbit-averaged torques between the stars moving on slowly-evolving elliptical orbits, drive a fast stochastic evolution of the stars' inclinations and eccentricities \citep{1996NewA....1..149R, 2018ApJ...860L..23B}. This is known as Resonant Relaxation. Importantly, the orbit-averaged dynamics leaves the semimajor axes unchanged. As argued in the original discovery paper, the relaxed state should be in a statistical  equilibrium, with the probability distribution function  proportional to the appropriate Boltzmann weight:
\begin{eqnarray}
    P(m, a, l, l_z,\omega,\Omega)&=&N(m, a)\times\nonumber\\
    & &\exp\left[-m\left(\beta\epsilon-\vec{\gamma}\cdot\vec{l}~\right)\right].
    \label{Prob}
\end{eqnarray}
Here $m$ is the mass and $a$ is the semimajor axis. On the left-hand side of the equation, the two Delaunay actions $l, l_z$ are the magnitude and the z-component of the specific angular momentum vector $\vec{l}$, and the two Delaunay angles $\omega, \Omega$ are the argument of periapsis and the longitude of ascending node.  On the right-hand side $N(m,a)$ is the normalization factor that will play no role in the following discussion,  and $\epsilon$ is the specific orbit-averaged potential energy of the star's interaction with the other stars in the cluster. The inverse temperature $\beta$ can be either positive, zero or negative, and $\vec{\gamma}$ is the rotational vector of the cluster; $\vec{\Omega}_{\rm cl}\equiv \vec{\gamma}/\beta$ is known as a thermodynamic angular velocity of the cluster, and it coincides with the angular velocity of the cluster's precession if the cluster is lopsided (e.g., Gruzinov et al.~2020). We can define the dimensionless rotation rate of the cluster as follows:
\begin{equation}
    \tilde{\gamma}=m_0 l_0 \gamma,
    \label{tildegamma}
\end{equation}
where $m_0$ is the characteristic mass of a star in the cluster and $l_0$ is the characteristic specific angular momentum of a stellar orbit in the cluster. We choose $m_0=M_\odot$ and $l_0=\sqrt{GMr_0}$, where $M$ is the mass of the supermassive black hole and $r=0.1$pc, the characteristic radius of the stellar disc. It is this rotation parameter that labels the plots in Figure 1.

A number of authors have explored thermodynamic equilibria of this form \citep{1996NewA....1..149R,2014JPhA...47C2001T, 2015MNRAS.448.3265K, 2017ApJ...842...90R, 2018PhRvL.121j1101S, 2019PhRvL.123b1103T, 2020MNRAS.493.2632T, 2020ApJ...905...11G, 2022MNRAS.514.3452M, 2022arXiv220207665M}.
We now show that for a slowly-rotating edge-on cluster, Eq.~(\ref{Prob}) together with some reasonably natural assumptions leads to a simple relationship
between the mean radial velocity and its dispersion, both measured from, e.g., pixel-integrated spectroscopy. This relationship can be used to estimate the rotational velocity of the relaxed cluster (or infer it, if the data is detailed enough).

Consider a cluster whose axis of rotation $z$ lies in the plane of the sky; thus $\vec{\gamma}=\gamma \hat{z}$. We shall assume that the cluster has a rotational symmetry about this axis. The Milky Way nuclear star cluster's rotation axis is perpendicular to the galaxy \citep{2008A&A...492..419T,2014A&A...570A...2F}, and the cluster appears to be axially symmetric in its inner parts despite some inferred triaxiality in its outer parts \citep{2017MNRAS.466.4040F}.  Let the $x$-axis be directed away from the observer along line of sight to the cluster, and let the origin of the $x, y, z$ coordinate system coincide with the center of the cluster. Each line of sight is characterized by coordinates $y,z$. The mean line-of-sight radial velocity is given by
\begin{equation}
    \langle v_r(y,z)\rangle=\int  P(\vec{r}, \vec{v})~v_x~dv_x~dv_y~dv_z~dx,
    \label{radvel}
\end{equation}
where $P(\vec{r}, \vec{v})$ is the probability distribution function for a star moving with velocity $\vec{v}$ to be located at position $\vec{r}$. The latter can be obtained by using Eq.~(\ref{Prob}) and expressing all the Delaunay variables in terms of $\vec{r}, \vec{v}$ (this is true because the Jacobian of a canonical transformation equals $1$).  Consider a reflection 
\begin{eqnarray}
    x&\rightarrow&-x\nonumber\\
    v_x&\rightarrow&-v_x\label{symmetry}
\end{eqnarray}
about the plane of the sky. This leaves $a,l,\epsilon$ unchanged, but flips the sign of $\vec{\gamma} \cdot \vec{l}$. Therefore,
\begin{eqnarray}
P(-x,y,z,-v_x, v_y, v_z)&=&\exp\left[-2 m\gamma(y v_x-x v_y)\right]\times\nonumber\\
  & &P(x,y,z,v_x,v_y, v_z).\label{flip1}
\end{eqnarray}
We assume that the cluster is rotating slowly, with $m \gamma l\ll 1$.
We therefore expand $\exp\left[-2 m\gamma(y v_x-x v_y)\right]\simeq 1-2 m\gamma(y v_x-x v_y)$. Multiplying the above equation by $v_x$ and integrating over the velocities and over $x$, we obtain the following relation:
\begin{equation}
    \langle v_r \rangle -m \gamma y \langle v_r \rangle^2=m\gamma \left(y\sigma_r^2-\langle x T_{xy}\rangle\right).
    \label{relation1}
\end{equation}
Here $\sigma$ is the radial velocity dispersion, $T_{ij}=\overline{v_i v_j}$ is the velocity tensor, and $\langle \rangle$ stands for the average along the line of sight. Solving for ${\gamma}$ and using Eq.~(\ref{tildegamma}), and assuming that $m=m_0$ is the typical mass of the observed stars, we get
\begin{equation}
    \tilde{\gamma}={l_0 \langle v_r\rangle\over y(\sigma_r^2+\langle v_r\rangle^2)-\langle x T_{xy}\rangle}.
\end{equation}
In the above equation everything is measurable except $\langle x T_{xy}\rangle$, since even if $v_y$ for individual stars could be measured, we would have no information about  $x$. Moreover, this term is in fact non-zero for anisotropic velocity ellipsoid and could be similar to $y \sigma_r^2$ in magnitude. Note however, $T_{xy}=0$ for an isotropic velocity distribution, which is expected to be produced by scalar resonant relaxation near the black hole\footnote{The prediction from the Resonant Relaxation is that the closer to the black hole, the more isotropic is the velocity ellipsoid.}. We shall assume this; in principle this assumption could be tested for consistency by checking that so-measured $\tilde{\gamma}$ is pixel-independent\footnote{\cite{2017MNRAS.466.4040F} performed orbit-modelling of the whole nuclear cluster out to $\sim 8$pc, and found that the anisotropy of the velocity ellipsoid is significant in the intermediate range of radii around $\sim 1$pc, but smaller at greater or much smaller radii. The inference is clearly not very precise in the regions where the stellar disc is located.}. We expect the result thus obtained to be correct to within a factor of $\sim 2$.

The data in \cite{2014A&A...570A...2F} is quite noisy at distances of interest. From Figure $11$ of that paper, the inner bin at $10$ arcsec (which corresponds to $\sim 0.4$pc) has $\langle v_r\rangle\sim 25 $km/sec and $\sigma\sim 85$km/sec. Plugging the numbers in the above equation, we get 

\begin{equation}\tilde{\gamma}\sim 0.3\end{equation}. 

We emphasize that this number is only an order of magnitude estimate and therefore we explore a range of values, as indicated in Figure $1$.

 \section{Analytical estimate of Resonant Friction timescale}
Resonant friction was first discussed as a phenomenon by \cite{1996NewA....1..149R}, as a dissipative counterpart to the stochastic resonant relaxation. Both are necessary for the thermal equilibrium to be established. The origin of the friction can be understood as follows. Consider the probability distribution in Eq.~(\ref{Prob}) for high masses, $m\gg m_0$. A star with such mass will tend to be near the orbit that minimizes the Jacoby constant $\epsilon-\vec{l}\cdot\vec{\gamma}/\beta$. \cite{2020ApJ...905...11G} studied such orbits and showed that they are stationary in the frame of reference that is rotating with the angular velocity $\vec{\gamma}/\beta$, and are typically aligned with the cluster's rotation. This occurs despite stochastic Vector Resonant Relaxation torques that are perturbing the massive orbit, because Resonant Friction {\it drives} massive objects in a cluster towards these special orbits. The existence of Resonant Friction is thus 
closely connected to the existence of thermodynamical equilibrium.

Resonant Friction explains why in numerical simulations of an Intermediate-Mass Black Hole inspiraling through a nuclear cluster, the former's orbit rapidly orients itself with the cluster's rotation. \cite{2012ApJ...754...42M} demonstrated explicitly in numerical experiments that this reorientation occurs because of secular torques, but they did not make the connection to thermodynamics (the community has been somewhat reluctant to accept Madigan \& Levin's arguments, instead attributing the reorientation to the $2$-body scattering processes).

It is the balance between the fluctuations and the dissipation that establishes the Boltzmann distribution. Below we use this fact to estimate the dissipation timescale in a rotating cluster.
First, we observe that the $z$-component of the angular momentum of an object inside the cluster experiences a. stochastic walk, due
to random torques from other orbits, and b. systematic drift upwards, which tends to align the orbit with the
cluster rotation. The corresponding evolution equation
for the $l_z$-distribution $f(m,l_z,t)$ can be written as
\begin{equation}
    {\partial f\over \partial t}=-{\partial\over \partial l_z}\left[F_{\rm drift}+F_{\rm stochastic}\right].
    \label{evolution}
\end{equation}

Here $m$ is the mass of a star, and $l_z=L_z/m$ where $L_z$ is the $z$-component of its angular momentum, and $F_{\rm drift}$ and $F_{\rm stochastic}$ are the fluxes in $l_z$-space due to the resonant friction and stochastic resonant relaxation, respectively.
The Fokker-Planck form of the fluxes is given by
\begin{eqnarray}
   F_{\rm drift}(m,l_z)&=&V(m,l_z)~f(m,l_z)\nonumber\\
   F_{\rm stochastic}(m,l_z)&=&-{\partial\over\partial l_z}\left[ D(l_z) f(m,l_z)\right]\label{fluxes}
\end{eqnarray}
The drift velocity $V(m,l_z)$ is mass-dependent, while the stochastic diffusion coefficient $D(l_z)$ is mass-independent because of the equivalence principle (the torque per mass on the orbit from the other stars depends only on the orbit, and not on the mass of the star).
Moreover, since flipping the direction of the angular momentum $\vec{l}\rightarrow-\vec{l}$ does not charge the period-averaged orbit,
we must have $D(l_z)=D(-l_z)$.

For slowly rotating, nearly spherically symmetric clusters $\epsilon$ depends only weakly on the orbit's orientation. Therefore,  the equilibrium distribution is given by
\begin{equation}
    f(m,l_z)=f_0(m)\exp(m\gamma l_z),
    \label{equilibrium}
\end{equation}
where $\vec{\gamma}=\gamma~\hat{z}$ is the rotational vector of the cluster. In equilibrium, $F_{\rm drift}+F_{\rm stochastic}=0$. Therefore,
substituting Eq.~(\ref{equilibrium}) into Eq~(\ref{fluxes}), we obtain the relationship for the  drift velocity
\begin{equation}
    V(m,l_z)=V(0,l_z)+m\gamma D(l_z).
\end{equation}
The second term of the right-hand side is the resonant-friction induced
part of the drift velocity. The first term
\begin{equation}
    V(0,l_z)=D^{\prime}(l_z)
\end{equation}
is the drift velocity of the zero-mass particles, required to enforce their
fully isotropic equilibrium distribution. The parity symmetry of $D(l_z)$ results in $V(0,0)=0$, and thus
\begin{equation}
    V(m,0)=m\gamma D(0),
\end{equation}
which establishes a key relationship between $V$, $\gamma$, and $m$.
It allows us to relate the frictional timescale $t_{\rm fr}$ to that of the vector resonant relaxation, $t_{\rm VRR}$. We have
\begin{eqnarray}
   t_{\rm fr}&\sim & l_0/V\sim l/(m\gamma D),\nonumber\\
   t_{\rm VRR}&\sim& l_0^2/D,\label{timescales}
\end{eqnarray}
where like in the previous section, $l_0$ is the characteristic specific angular momentum of an orbit in the cluster. Thus we have
\begin{equation}
    t_{\rm fr}\sim \tilde{\gamma}^{-1} {m_0\over m} t_{\rm VRR},
    \label{tfr1}
\end{equation}
where $m_0$ is the mass of a typical star in the cluster and like before, the the dimensionless rotation parameter is given by
\begin{equation}
    \tilde{\gamma}=m_0\gamma l_0.
\end{equation}

So what is $t_{\rm VRR}$? Following the arguments of \cite{1996NewA....1..149R}, one can show that
\begin{equation}
    t_{\rm VRR}\sim {M_{\rm BH}^2\over M_* m_0} {P^2\over t_{\rm coh}},
\end{equation}
where $M_{\rm BH}$  is the mass of the central black hole, $M_*$ is the mass of the stellar cluster, $P$ is the characteristic orbital period, and  $t_{\rm coh}$ is the characteristic coherence timescale for the fluctuating torque. \cite{1996NewA....1..149R} took $t_{\rm coh}\sim t_{\rm VRR}$, arguing that VRR is the main mechanism for the change in orientation of the orbits. Thus they obtain
\begin{equation}
    t_{\rm VRR}\sim {M_{\rm BH}\over \sqrt{M_* m_0}}P.
    \label{tVRR1}
\end{equation}
Substituting this into Eq.~(\ref{tfr1}), we obtain
\begin{equation}
    t_{\rm fr}\sim {M_{\rm BH}~P\over \tilde{\gamma}\sqrt{M_* m_0}}.
    \label{tfr2}
\end{equation}
However, this argument has a serious limitation. If a cluster is rotating, its stellar distribution is flattened, with the ellipticity $\sim \tilde{\gamma}^2$. This ellipticity drives precession of the stellar orbits with characteristic timescale
\begin{equation}
    t_{\rm prec}\sim {M_{\rm BH}\over M_*}\tilde{\gamma}^{-2}P.
\end{equation}
This timescale becomes comparable the one given in Eq.~(\ref{tVRR1}) for $\tilde{\gamma}\sim N_s^{-1/4}\sim 0.1$, where $N_s=M_*/m_0$ is the number of stars in the cluster. Therefore for $\tilde{\gamma}\gg N_s^{-1/4}$, we should consider $t_{\rm coh}\sim t_{\rm prec}$. In that case,   we get
\begin{equation}
t_{\rm VRR}\sim {M_{\rm BH}\over m_0}\tilde{\gamma}^2 P.
\end{equation}
Substituting this into Eq.~(\ref{tfr1}), we get
\begin{equation}
t_{\rm fr}\sim \tilde{\gamma} {M\over m} P.
\label{tfr3}
\end{equation}
Note that this does not depend on the cluster mass $M_*$. To sum up: if $\tilde{\gamma}\ll N_s^{-1/4}$, one should use Eq.~(\ref{tfr2}); in the opposite case, one should use Eq.~(\ref{tfr3}). Clearly, these expressions are very approximate, and one needs more precise arguments which are beyond the scope of this paper,
to obtain more reliable expressions with more precisely stated domains of validity. 

Having derived the Resonant Friction timescale for a stellar orbit of mass $m$, we note that the effect should be there for any massive object inside the cluster that is gravitationally coherent. Thus it should apply to a disc of stars, for as long as the orbital planes of the stellar orbits remain clustered. We thus apply the expression above to the whole  stellar disc. Taking $m=M_{\rm disc}=8000 M_\odot$, $M_{\rm BH}=4\times 10^6 M_\odot$, $\tilde{\gamma}=0.3$, and $P=1500$yr (corresponding to a circular orbit at $0.1$pc), we get 
$t_{\rm fr}\sim 2.5$Myrs. This is consistent with the results of our numerical experiments described in Section 2.

\section{Alignment of accretion discs with the nuclear cluster rotation}
\begin{figure*}[t]
   \centering
    \epsfxsize=16cm 
    \epsfbox{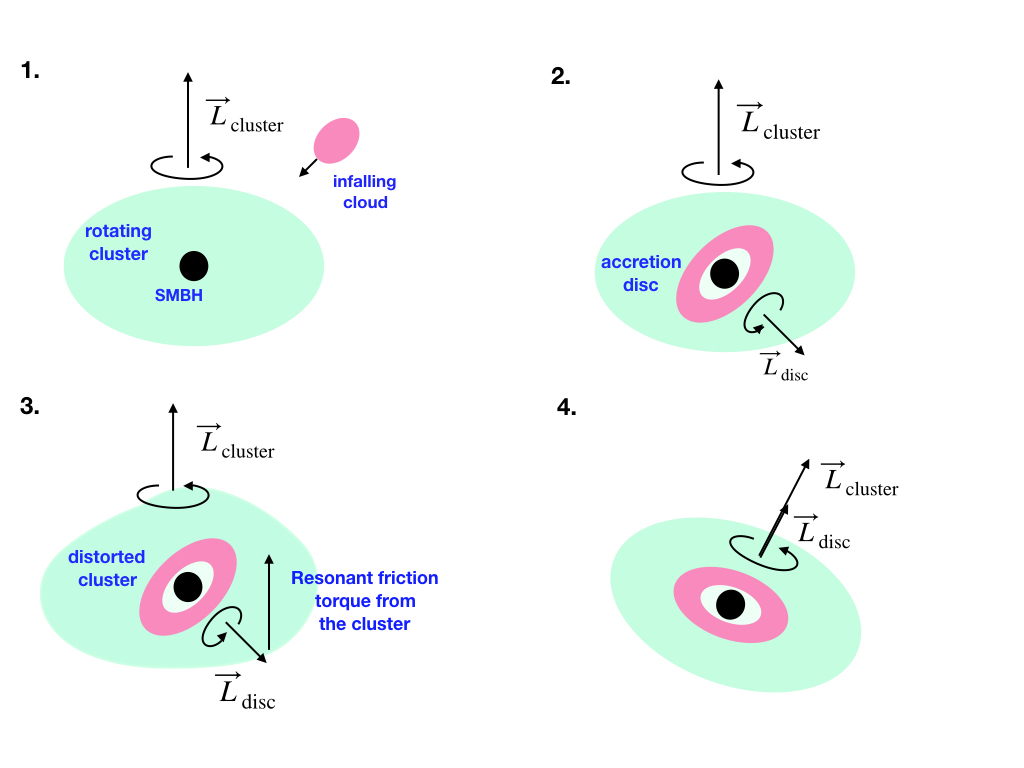} 
 \caption{
 \small
Pictorial description of the scenario described here. An infalling cloud of gas is tidally disrupted and its material forms an accretion disc around the SMBH. The initial orientation of the disc is random, but the Resonant Dynamical Friction drives the disc's angular momentum $\vec{L}_{\rm disc}$ into alignment with that of the cluster, $\vec{L}_{\rm cluster}$.  }
   \end{figure*} 
 
\subsection{General remarks}
A classic argument by \cite{1982MNRAS.200..115S}, 
and its more recent elaborations 
\citep{2002MNRAS.335..965Y, 2012AdAst2012E...7K}, 
strongly suggest that supermassive black holes 
(SMBH) 
acquire most of their mass by accreting gas from thin discs in galactic nuclei. If the orientation of an accretion disc
relative to the black hole could be maintained, this mode of accretion would drive the black hole to a  very high spin, with dimensionless spin parameter $\alpha>0.9$  
\citep{1973blho.conf..343N}. 
The discovery of 
\cite{1975ApJ...195L..65B} 
that gravito-magnetic forces drive an accretion
disc into alignment with the equatorial plane near the black hole, would suggest that we could expect essentially all supermassive black holes to be rapidly spinning. Measurements  of  x-ray  and  radio  emission  from accreting supermassive black holes in galactic nuclei indicate that they are indeed rotating rapidly \citep{2020arXiv200808588J,2019ApJ...886...37D}. The spin energy of SMBHs is thought to be responsible for powering relativistic jets emanating from galactic nuclei, and  thus SMBH spin is the key agent of feedback during the formation of elliptical galaxies. 

However, much remains to be understood about the spins of SMBHs. The measurements of the black hole spins rely on rather complex models of accretion discs and jets, and thus may contain systematic uncertainties. On the theoretical side, 
\cite{2006MNRAS.373L..90K} 
argued that the accretion is likely to be driven by randomly-oriented infall episodes; 
this is supported observationally by the fact that many radio jets seem to be randomly oriented relative to their host spiral galaxies. In this stochastic-accretion picture, one expects  nearly half of the transient accretion discs to be counter-rotating relative to the black hole. 
Importantly, a counter-rotating innermost stable orbit has a greater radius than a co-rotating innermost stable orbit, and therefore the counter-rotating discs have a larger lever arm. \cite{2008MNRAS.385.1621K} argued that as a result  of this, the supermassive black holes are spun down, on average, to low spins of $\alpha\sim 0.3$. A more elaborated model of \cite{2018MNRAS.477.3807F} makes a distinction between lower-mass black holes of $M_{\rm BH}\lesssim 10^7M_\odot$ and the more massive ones. These authors argue that despite of the stochastic feeding, the spin directions of the lower mass black holes are expected to remain adiabatically aligned with the
angular momentum directions of their accretion discs, while the higher-mass black holes might experience both spin-ups and spin-downs in equal measure.  

LISA will provide exquisitely precise measurements of the supermassive black hole spins, by measuring the gravitational waves from mergers of  stellar-mass and intermediate-mass black holes with SMBHs. It is thus imperative to develop reliable theoretical predictions for the SMBH spins.
Below we argue that a key piece of physics is missing in the current theoretical analyses, namely the stabilizing effect of rotating black hole clusters on the orientation of the accretion disc. 
Indeed if the stellar disc in our Galactic Center gets reoriented into alignment with the cluster, why would not an accretion disc in a galactic nucleus which is more active than ours?

\subsection{Resonant friction on accretion discs}
The scenario we consider is depicted in Figure 2. In its first stage, a cloud of gas, with the mass of $\sim 10^4M_\odot$ falls in, gets tidally disrupted and forms a gaseous accretion disc. 
If the cluster is  flattened due to its rotation, it will  exert a gravitational torque on the
disc, causing it to precess. The differential precession could distort the disc. The hydrodynamics of distorted discs could be complicated, but it is reasonable to expect that after much dissipation from shocks etc., the disc would reassemble in the equatorial plane of the cluster. Naively it would seem that the disc could equally likely be co-rotating and counter-rotating with the cluster.

However, the resonant friction is expected to break the symmetry between co- and counter-rotation, and drive the disc's and the cluster's angular momenta into alignment. 
Resonant friction should act on any massive object inside the cluster, in particular it should affect the accretion disc, so long as the disc remains coherent and creates a gravitational perturbation that affects the cluster. Whether the disc remains coherent long enough to experience the effect is the key question.
In absence of a full hydrodynamical treatment, our intuition can be guided by simulations of stellar disc evolution inside a rotating cluster, such as the one illustrated in Figure $1$. 
The initial disc remains coherent enough, until its overall orbital angular momentum flips into co-rotation with the cluster.  If instead of the stellar disc we introduced a gaseous disc of the same mass, it would
firstly flip into co-rotation with the cluster, and then settle into the equatorial plane of the cluster due to the hydrodynamic torques that are caused by the disc's differential precession.

If this picture were correct (and it obviously needs to be tested by hydrodynamic simulations!), then the rotating clusters could serve as
a stabilizing flywheel for the material that accretes onto the SMBH. It would imply that
\begin{itemize}
    \item The SMBH spin direction is aligned with that of its host cluster
    \item The SMBH spin magnitude could reach a very high value, unless there exists an as yet undetermined mechanism for the SMBH spindown (such as its super-radiant coupling to a scalar field or a cosmic string)
    \item Since the resonant friction from cluster rotation will affect inspiraling Intermediate-Mass 
    Black Holes, one expects a non-trivial alignment between the inspiraling orbit and the SMBH spin. Such an alignment will be  detectable by LISA.
\end{itemize}

In proposing hydrodynamic numerical experiments, we need to ascertain that the scenario is reasonable by comparing the resonant friction timescale with that of the dissolution of the disc due to a differential precession. 
As argued in the previous section, the timescale for the friction to align the disc is
\begin{equation}
    t_{\rm fr}\sim  \tilde{\gamma} ~{M_{\rm BH}\over M_{\rm disc}}~P,
\end{equation}
this equation is valid so long as $\tilde{\gamma}>N_s^{-1/4}$, where $N_s$ is the number of stars in the cluster. Note that this timescale is comparable to the period of a bending mode of the disc due to its self-gravity, $t_{\rm bend}\sim (M_{\rm BH}/M_{\rm disc}) P$. Therefore the reorientation of the disc due to Resonant Friction is expected to be accompanied by  excitation of the bending motion of the disc. This explains the excitation of the
disc warps,  and the disc's breaking into several rings in some of our simulations.

Note also that the mass of the cluster, assumed to satisfy $M_{\rm disc}<M_{*}<M_{\rm BH}$, does not enter into the expression above; this is because the cluster contributes to the timescale in two distinct ways, and the contributions cancel each other. On the one hand, the more stars are affected by the disc, the stronger is their gravitational back-reaction on the disc. On the other hand, the rate of precession of the orbital angular momenta of the cluster
stars is  $\propto M_*$, and a fast rate of precession is suppressing the cluster's gravitational response to the disc. The rate of differential precession is given by
\begin{equation}
    t_{\rm prec}\sim \tilde{\gamma}^{-2} ~ (M_{\rm BH}/M_{*})~P.
\end{equation}
Comparing the two timescales,

\begin{equation}
    {t_{\rm fr}\over t_{\rm prec}}\sim 0.5 \left({\tilde{\gamma}\over 0.3}\right)^3~{M_{*}\over 15 M_{\rm disc}}.
\end{equation}
\newline
In the numerical example above, we used the parameters relevant for the Galactic Center stellar disc, with $M_*$ being the mass of the stars within the disc's outer radius.
The estimate suggests that for a realistic parameter range the disc alignment is promising. Neglected in this estimate is are the effects of Reynolds stresses,  which are expected to enhance the disc's coherence and thus increase the robustness of alignment. We thus conclude that the idea deserves further theoretical investigation,  with numerical experiments that involve both stellar and gas dynamics. 


\section{Discussion}
Rotating nuclear star clusters in galactic nuclei are giant cosmic centrifuges. Any heavy object inside the cluster (an IMBH or a massive stellar or gas disc) feels a very strong Resonant Friction torque that tends to align its orbit with the cluster's rotation. This torque can (partially) disrupt an impulsively injected disc, and we have given strong arguments that this is what happened with the young stellar disc at the center of the Milky Way. We speculate that this process is also efficient for stochastically injected gaseous accretion discs. If future numerical experiments confirm our speculation, then there would be  reason to believe that the direction of spin of the supermassive  black hole could be anchored to that of its host cluster. Contrary to the arguments of \cite{2008MNRAS.385.1621K}, the episodes of successive accretion would then allow the black hole to acquire high rotation rate. We must note, however, that the SgrA* shadow observations by the EHT collaborations, disfavour the edge-on spin of the supermassive black hole \citep{2022ApJ...930L..12E}. These interesting, albeit model-dependent measurements need to be considered as the models of the disc-cluster interaction become more reliable.

The result for the Milky Way's young stellar disc has special significance. Its partially disrupted state has been a puzzle since its discovery, and several authors including the current one appealed to massive perturbers to produce the necessary gravitational violence \citep{2005ApJ...635..341L, 2020ApJ...905..169Z, 2021ApJ...919..140S}. This paper demonstrated that hypothetical massive perturbers are unnecessary for explaining the observations.

This work has several technical shortcomings that will need to be improved upon in future work. In our simulations we considered only low-order multipole interactions, which are likely adequate for the study of the long-range torques responsible for resonant friction, but perhaps underestimate the stochastic torques experienced by the stars, which may negatively impact the coherent parts of the inner disc. Definitive confirmation should come from a direct (and extremely expensive) $N$=body simulations, similar to the ones performed in \cite{2022MNRAS.tmp.2848P}, but modified to include  rotation of the host cluster. Resonant friction in rotating clusters deserves a more rigorous analytical treatment than the scaling arguments given here. Gas dynamics of accretion discs warped by resonant friction torques is an interesting topic of future study.

We thank Andrei Gruzinov and Scott Tremaine for insightful discussions, and for raising important conceptual and practical points after reading the first draft of this paper. We thank Sophie Koudmani for explaining the current theoretical models for the black-hole spin evolution, and Sarah Levin for advice on the prose. This research is supported by the Simons Investigator Grant 827103.

\appendix
\section{Numerical scheme for quadrupole-driven evolution}
As explained in the text, we represent the stars as circular rings with mass $m_i$ and radii $r_i$.  In only the quadrupolar part of the potential of the rings is taken into account, then the torque acting on the $i$'th ring 
is given by (see e.g., Kocsis \& Tremaine 2015):
\begin{equation}
    \vec{\tau}_{i}=G m_i \vec{n}_i\times\sum_j a_{ij} m_j\overleftrightarrow{P}_j ~\vec{n}_i.
\end{equation}
Here 
\begin{equation}
    a_{ij}={3\over 4} {\min(r_i,r_j)^2\over \max(r_i, r_j)^3},
\end{equation}
and
$\overleftrightarrow{P}_j$ is a projection operator, with $\overleftrightarrow{P}_j \vec{k}\equiv (\vec{k}\cdot\vec{n}_j)\vec{n}_j$.
The dimensionless equations of motion are given by
\begin{equation}
    {d \vec{n}_i\over dT}=\vec{n}_i\times \overleftrightarrow{Q}_i~\vec{n}_i, 
    \label{motion1}
\end{equation}
where 
\begin{equation}
    T={\sqrt{GM\over r_0^3}}{m_0\over M}t,
\end{equation}
and $\overleftrightarrow{Q}_i$ is a $3\times 3$ matrix given by
\begin{equation}
    \overleftrightarrow{Q}_i=\sum_j A_{ij}M_j \overleftrightarrow{P}_j.
\end{equation}
Here $r_0$ and $m_0$ are some convenient reference radius and mass, taken to be $0.1$pc and $M_\odot$ for our purposes.  Numerically we have  $T\simeq (t/10^9\hbox{yr})$. Furthermore, $M_i\equiv m_i/m_0$, and $A_{ij}\equiv r_0 a_{ij}$.

The evaluation of each $ \overleftrightarrow{Q}_i$ requires $\propto N_s$ operations, where $N_s$ is the number of stars. So each timestep requires, in general, $\propto N_s^2$ operations. Let us assume now that the stars are binned into $N_g\ll N_s$ groups, with the orbits of stars that belong to the same group assumed to have the same radius. Let's label the groups with Greek letters $\alpha, \beta, ...$ and let $N_\alpha$ be the number of stars in group $\alpha$. Furthermore let us label the stars within each group by a Latin letter, so $\vec{n}_{\alpha i}$ stands for the vector characterizing the $i$'th star in group $\alpha$.  Let us define 
\begin{equation}
    \hat{P}_\alpha=\sum_{i=1}^{N_\alpha} M_{\alpha i}\overleftrightarrow{P}_{\alpha i},
\end{equation}
and
\begin{equation}
    \hat{Q}_\alpha=\sum_\beta A_{\alpha\beta}\overleftrightarrow{P}_\alpha.
\end{equation}

The equations of motion are then
\begin{equation}
{d\vec{n}_{\alpha i}\over dT}=\vec{n}_{\alpha i}\times \hat{Q}_\alpha \vec{n}_{\alpha i}.
\label{motionmeanfield}
\end{equation}
It takes $\propto N_s=\sum_\alpha N_\alpha$ operations to compute all matrices $\hat{P}_\alpha$ and another $\propto N_g^2$ operations to compute all the matrices $\hat{Q}_\beta$. Thus for $N_g\lesssim \sqrt{N_s}$, the computations are extremely fast and not sensitive to the number of bins.

The implementation of this algorithm in Matlab can be extremely fast since all aspects of the computation can be ``vectorized'' and ``matricized''. For this purpose it is convenient to arrange the number of the stars in each bin to be the same, and represent their $\vec{n}$-vectors as parts of a $3$-dimensional array where one dimension marks a bin number, the second one identifies a star inside the bin, and the third dimension marks the $3$ components of the vector.

\section{Higher orders}
The  benefits of radial binning can be extended to higher-order potential terms, so long as the coupling coefficients between different orbits are a function of their radii only. A general potential energy of interaction between the orbits $i$ and $j$ can be expanded in a series
\begin{equation}
    U_{ij}=m_i m_j\sum_{k=0}^{\infty} u_k(r_i,r_j) \left(\vec{n}_i\cdot\vec{n}_j\right)^{2k}.
    \label{B1}
\end{equation}
The torque on the $i$'th star exerted bu the $j$'th star is give by
\begin{equation}
    \vec{\tau}_{ij}=-\vec{n}_i\times {\partial U_{ij}\over \partial \vec{n}_i}.
    \label{B2}
\end{equation}
Let $i$ be the label for a star that belongs to the bin $\alpha$ with the radius $r_\alpha$. Let's determine the torque on this star, by summing contributions from  from all the stars in all bins $\beta$ (this includes the possibility $\beta=\alpha$). We define $u_{k\alpha\beta}\equiv u_k(r_\alpha,r_\beta)$, and obtain:
\begin{equation}
    (1/m_{i})\vec{\tau}_{i}=-\vec{n}_{i}\times \sum_k 2k\vec{W}_{(k\alpha)}(\vec{n}_i).
    \label{B3}
\end{equation}
The components of $\vec{W}_{(k\beta)}$ are given by
\begin{equation}
     W_{(k\alpha)p_1}= \sum_{p_2 p_3 ... p_k}w_{(k\alpha)p_1 p_2...p_k}n_{ip_2} n_{ip_3}...n_{i p_{k}}.
\end{equation}
where 
\begin{equation}
    w_{(k\alpha)p_1 p_2 ... p_k}=\sum_\beta u_{k\alpha\beta}\sum_{j(r_j=r_\beta)} m_{j} n_{jp_1}n_{jp_2}...n_{jp_k}.
    \label{B5}
\end{equation}
In the equation above, $n_{jp}$ stands of the $p$-component of $\vec{n}_j$, and the second sum on the right-hand side is evaluated over all the stars $j$ in the bin $\beta$.

The advantage of this approach is that one only needs to evaluate tensors $w_{(k\beta)}$ each timestep. Just like in the quadrupole case, the number of evaluations one needs to make scales as $N_s$ instead of 
the more common $N_s^2$ for $N_b\ll \sqrt{N_s}$. In detriment to this way of doing the dynamics, the 
operations with high-ranking tensors become exponentially more expensive as the rank of the tensor increases. Luckily, for the purpose of this paper we see that there are only modest difference between the simulations in which we include $2$ or $3$ terms in the Eq.~(\ref{B2}), so we therefore limit ourselves to $3$-rd order terms.

In practice we proceed as follows. For a fixed angle $\theta$ between $\vec{n}_i$ and $\vec{n}_j$, the interaction energy can be written as an expansion in 
\begin{equation}
    \alpha_{ij}\equiv {2 r_i r_j\over r_i^2+r_j^2}.
\end{equation}
Expanding up to the fourth order and collecting the relevant terms at $(\vec{n}_i\cdot\vec{n}_j)^2$ and $(\vec{n}_i\cdot\vec{n}_j)^4$, we get up to an additive constant
\begin{equation}
   U_{ij}=-{G m_i m_i\over\sqrt{r_i^2+r_j^2}}\left[{3\over 32}\alpha_{ij}^2\left(1+{105\over 384}\alpha_{ij}^2\right) (\vec{n}_i\cdot\vec{n}_j)^2+{315\over 8192}\alpha_{ij}^4 (\vec{n}_i\cdot\vec{n}_j)^4\right].
   \label{secondorder}
\end{equation}
This is the expression we are using instead of the full expansion in Eq.~(\ref{B1}).

\bibliography{ms4}

\clearpage

\end{document}